\documentclass[12pt]{iopart}

\usepackage{iopams}  
\usepackage{graphicx} 
\usepackage{xcolor}

\expandafter\let\csname equation*\endcsname\relax
\expandafter\let\csname endequation*\endcsname\relax

\usepackage{hyperref}
\usepackage{amsmath}

\begin{document}

\title{Impact of Hydrogenation on the Stability and Mechanical Properties of Amorphous Boron Nitride}

\author{Onurcan Kaya$^{1,2}$, Luigi Colombo$^3$, Aleandro Antidormi$^1$, Marco A. Villena$^4$, Mario Lanza$^4$, Ivan Cole$^{2}$ and Stephan Roche$^{1,5}$}

\address{Catalan Institute of Nanoscience and Nanotechnology (ICN2), CSIC and BIST,
Campus UAB, Bellaterra, 08193, Barcelona, Spain}
\address{School of Engineering, RMIT University, Melbourne, Victoria, 3001, Australia
}
\address{Department of Materials Science and Engineering, The University of Texas at Dallas, Richardson, TX, 75080, USA}
\address{Department of Material Science and Engineering, King Abdullah University of
Science and Technology, Thuwal, 23955, Saudi Arabia.}
\address{ICREA Institucio Catalana de Recerca i Estudis Avancats, 08010 Barcelona, Spain}

\ead{stephan.roche@icn2.cat}

\begin{abstract}
Interconnect materials with ultralow dielectric constant, and good thermal and mechanical properties are crucial for the further miniaturization of electronic devices. Recently, it has been demonstrated that ultrathin amorphous boron nitride (aBN) films have a very low dielectric constant, high density (above 2.1 $\rm g/cm^3$), high thermal stability, and mechanical properties. The excellent properties of aBN derive from the nature and degree of disorder, which can be controlled at fabrication, allowing tuning of the physical properties for desired applications. Here, we report an improvement in the stability and mechanical properties of amorphous boron nitride upon moderate hydrogen content. With the introduction of a Gaussian approximation potential (GAP) for atomistic simulations, we investigate the changing morphology of amorphous boron nitride with varying H densities. We found that for 8 at\% of H doping, the concentration of $\rm sp^3$-hybridized atoms reaches a maximum which leads to an improvement of thermal stability and mechanical properties by 20\%.  These results will be a guideline for experimentalists and process engineers to tune the growth conditions of amorphous boron nitride films for numerous applications.

\end{abstract}

\section{Introduction}

As the semiconductor industry downscales integrated circuits and power consumption increases, materials and reliability are pushed to their limits. It is thus increasingly important to either improve the performances of existing materials and/or develop new materials to meet these stringent demands of power reduction of electronic devices and circuits. While transistors have gone through several generations of design and the introduction of new materials, the back-end-of-line interconnects have seen fewer changes. Nevertheless, significant efforts have been dedicated to address the reduction of the resistance-capacitance (RC) delay \cite{HOOFMAN2005337, Grill2009, Palumbo2020}. The resistance-capacitance (RC) reduction could be achieved in several ways, 1) decrease the capacitance density by decreasing the dielectric constant of interlayer dielectric (ILD) and the intermetal dielectric (IMD), 2) decrease the resistivity of the metal interconnect wiring, and 3) increase the cross-section of the interconnects. However, there are several problems associated with decreasing RC time delay as per the above approaches, 1) new materials with lower dielectric constants exist but may not be stable within the required process flow, that is they may not be mechanically and thermally stable as well as being good diffusion barriers, and 2) it is difficult to find metallic systems with electrical conductivity lower than Cu that can also meet the stability requirements of devices \cite{Grill2016, Noguchi2005, Palov2106}.\\
A potential new barrier dielectric has recently emerged, amorphous boron nitride ($\rm \alpha$-BN). Experimental reports on $\rm \alpha$-BN indicate that it has a low dielectric constant, k-values lower than 2, and exhibits higher stability and mechanical properties compared to other low dielectric materials such as organic polymeric materials \cite{Hong2020, Glavin2016, Lin2022}. In addition, theoretical predictions suggest that a certain density of carbon content improves the structural and thermal properties of $\rm \alpha$-BN:C \cite{Kaya2023}. More recently there is also a study on the use of $\rm \alpha$-BN for interconnect capping layers \cite{Kim2023}. This last work reports on the plasma-enhanced chemical vapor deposition (PECVD) of 3 and 7 nm  $\rm \alpha$-BN as a capping layer to replace PECVD-grown ${\rm Si_3N_4}$. The study finds that $\rm \alpha$-BN is an excellent insulator with efficient barrier against Cu diffusion, has good adhesion to copper and ${\rm SiO_2}$, is thermally stable and has a much lower dielectric constant (k=3), than ${\rm Si_3N_4}$ (k$\sim$7) enabling an RC-delay reduction of 10-17\%. In Ref.\cite{Kaya2023}, using machine learning techniques and classical molecular dynamics, we explored the effects of C content on the physical properties of $\rm \alpha$-BN in an attempt to create an even more stable dielectric \cite{Kaya2023}. However, neither these results nor the previous studies on ALD-grown $\rm \alpha$-BN\cite{FERGUSON200216, MARLID2002167}, report on the effects of C and H content on the properties of the films. Also, it is well known that PECVD-grown ${\rm Si_3N_4}$ can contain large amounts of hydrogen \cite{BENOIT20116550} and it is likely that PECVD-grown $\rm \alpha$-BN may also contain hydrogen. Recently, Jacquemin et al. \cite{Jacquemin2023} showed that in the absence of hydrogen atoms, bromide ions can block the formation of $\rm sp^2$-hybridized atoms and produce continuous and stable $\rm \alpha$-BN films in a PECVD reactor using the $\rm BBr_3$ and $\rm N_2$ precursor. Hong et al. \cite{Hong2020} measured 5.5\% of H contamination to their PECVD grown $\rm \alpha$-BN when they used borazine as a precursor. Nonetheless impact of H on the stability or mechanical properties of $\rm \alpha$-BN is not clear.
 Therefore, it is important to understand the stability of hydrogen in $\rm \alpha$-BN as it is for ${\rm Si_3N_4}$, since hydrogen can impact the underlying performance of the Si transistors and affect the dielectric and physical properties of interconnects. In this paper, we clarify the effects of hydrogenation of $\rm \alpha$-BN:H on its thermal stability and mechanical properties.  \\
From a computational perspective, atomistic calculations represent a suitable tool to describe complex structures, giving access to details at the atomic and molecular level to enable a basic understanding of new materials without performing more costly and time-consuming experiments. The extremely disordered nature of amorphous materials requires a computational approach able to capture the interatomic potentials in arbitrary complex local environments, a challenge that can only be tackled with machine learning-based methods \cite{Musil2021,Sivaraman2020,Schleder_2019}. More specifically, classical molecular dynamics with the employment of force fields derived using machine learning and ab-initio techniques constitute a powerful methodology to describe disordered material while keeping first-principles accuracy \cite{Deringer2017,Unruh2022,QIAN2019,Mortazavi_2020}. In our theoretical study, we found that hydrogen influences the hybridization of the core structure ($\rm sp^2/sp^3$) and increases structural disorder as observed through radial distribution function (RDF). Further, in comparison to carbon doped $\rm \alpha$-BN, whose thermal stability and Young’s modulus increase monotonically with C doping, hydrogen doping leads to an $\rm \alpha$-BN with a non-monotonic increase in these properties, in fact, they peak at around 8 at\% hydrogen. These results are critically important in providing directions to the experimentalist in tuning the deposition processes to meet the electronic device requirements. 

\section{Methods}
\subsection{Gaussian Approximation Potentials (GAP)}
\begin{figure}[ht]
\centering
    \includegraphics[width=1.0\columnwidth]{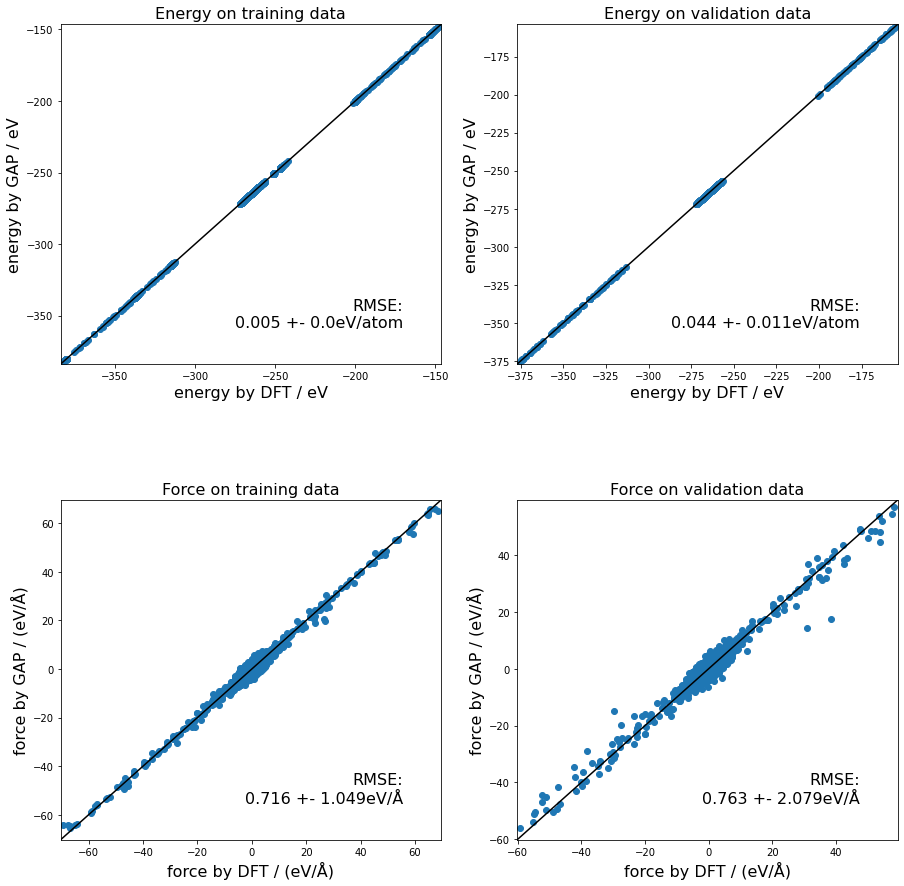}
    \caption{Scatter plots of the energies and forces obtained using newly generated GAP model of the corresponding DFT values computed over both training and validation datasets.}
    \label{fig:gap}
\end{figure}

Structures for training and validation sets are generated using DFT calculations using the Quantum Espresso package \cite{giannozzi2009quantum,giannozzi2017advanced,giannozzi2020quantum} with Perdew–Burke–Ernzerhof (PBE) \cite{perdew1996generalized} exchange-correlation functional and projector-augmented wave (PAW) pseudopotentials. The energy cutoff for the wavefunction and electronic density is 75 Ry and 600 Ry, respectively. Both training and validation sets contain sufficiently large data sets of forces, energies, and stresses. The mentioned data sets involve both crystalline and amorphous BN structures, $\rm \alpha$-BN:H samples with different levels of H concentration.  
The density of the structures in the data sets ranges between 1.0 $g/cm^3$ and 3.0 $g/cm^3$. Moreover, they also contain several distinct molecular configurations ($\rm H_2$, $\rm N_2$, ammonia, ammonium, and borazine) and isolated B, N, and H atoms. The final training and validation sets contain 2500 and 1800 samples, respectively. Such a wide variety of atomistic configurations enables us to model $\rm \alpha$-BN:H samples with better accuracy. \\   
The parameters shown in Table \ref{table:gap} has been employed to train the GAP model for $\rm \alpha$-BN:H using the training database. Smooth Overlap of Atomic Positions (SOAP) descriptor\cite{Bartok2013} is introduced to model the many-body interactions between atoms, while 2b and 3b descriptors are adopted for two-body and three-body interactions. The model is trained using the total energy function (Eq. 1) and local energy contributions, where 2b, 3b and MB represent the two- and three-body and many-body descriptors, where $\rm \delta^d$ is a scaling parameter which represents the contribution of each term, $\rm \epsilon$ is the local energy contribution, K is the kernel function, \textbf{q} is the training configuration and $\rm \alpha_t$ is the all fitting coefficients \cite{Deringer2017}. 
\begin{multline}
    E = (\delta^{2b})^2\sum_{i} \sum_{t} \alpha_t^{2b} K^{2b} (\textbf{q}_i^{2b} , \textbf{q}_t^{2b}) + (\delta^{3b})^2\sum_{j} \sum_{t} \alpha_t^{3b} K^{3b} (\textbf{q}_j^{3b} , \textbf{q}_t^{3b}) \\ + (\delta^{MB})^2\sum_{a} \sum_{t} \alpha_t^{MB} K^{MB} (\textbf{q}_a^{MB} , \textbf{q}_t^{MB}) 
\end{multline} 
Uniform sparsification has been used for 2b and 3b terms, while the CUR method\cite{mahoney2009cur} has been chosen for the SOAP kernel. After the training, we compare the energies and forces of structures in both training and validation sets obtained from molecular dynamics simulations with GAP model (GAP-MD) with the results from DFT calculations in order to evaluate the accuracy of the generated GAP model as shown in Fig.~\ref{fig:gap}. A significantly low root mean squared error (RMSE) is calculated for both training and validation sets. \\
\begin{table}
\begin{center}
    \begin{tabular}{cccc}
    \hline 
    & 2-body& 3-body& SOAP\\ \hline \hline
    $\delta$ (eV)& 2.0&0.1& 0.1\\ 
    $r_{cut}$ ($\AA$)& 3.7& 3.0&3.7\\
    $r_{\Delta}$ ($\AA$)& & & 0.5\\ \hline
    $\sigma_{at}$ ($\AA$)& & &0.5\\
    $n_{max}$, $l_{max}$ & & &8\\
    $\zeta$& & &4 \\ \hline
    Sparsification& Uniform & Uniform& CUR \\ \hline
    $N_t$ ($\rm \alpha$-BN bulk)& & 150&2000\\
    $N_t$ (Crystalline samples)& & 50&500\\
    $N_t$ (Total)& 15& 200&2500\\ \hline
    \end{tabular} 
    \label{table:gap}
    \caption{Parameters used to train the GAP potential for H-doped $\rm \alpha$-BN.}
\end{center}
\end{table}

\subsection{Melt-Quench Protocol for Sample Generation}
One of the most common strategies used in molecular dynamics (MD) simulations to generate amorphous samples is the melt-quenching protocol. In this method, the sample is first melted by heating above the melting temperature and then rapidly quenched \cite{SANTOS201920}. The $\rm \alpha$-BN:H samples containing varying amounts of hydrogen are generated following this protocol using GAP-MD simulations with Large-scale Atomic/Molecular Massively Parallel Simulator(LAMMPS) code\cite{LAMMPS}. Each sample has 10000 atoms and an equal number of boron and nitrogen atoms with varying amount of H. Each edge of cubic simulation boxes is ranging between 40 - 45 $\AA$ depending the H concentration to keep the initial density of samples same. First, all boron, nitrogen, and hydrogen atoms are placed randomly in the simulation cell; then, the melted samples are equilibrated at 5000 K for 50 ps (timestep of 0.1 fs) using a Nosé-Hover thermostat. Later, the temperature of the samples is reduced to 2000 K in 100 ps (with a cooling rate of 40 K/ps) and equilibrated at this temperature. Samples are then cooled down to 300 K in 150 ps. After a short relaxation (10 ps) run, we also applied an annealing step where the temperature was increased to 700 K with a heating rate of 20 K/s and decreased to 300 K with a cooling rate of 5 K/s. Finally, annealed samples are relaxed and equilibrated at 300 K for 50 ps in the NPT ensemble.\\
\section{Results}
\subsection{Morphology Analysis of $\rm \alpha$-BN:H}
\begin{figure}[ht]
\centering
    \includegraphics[width=0.83\columnwidth]{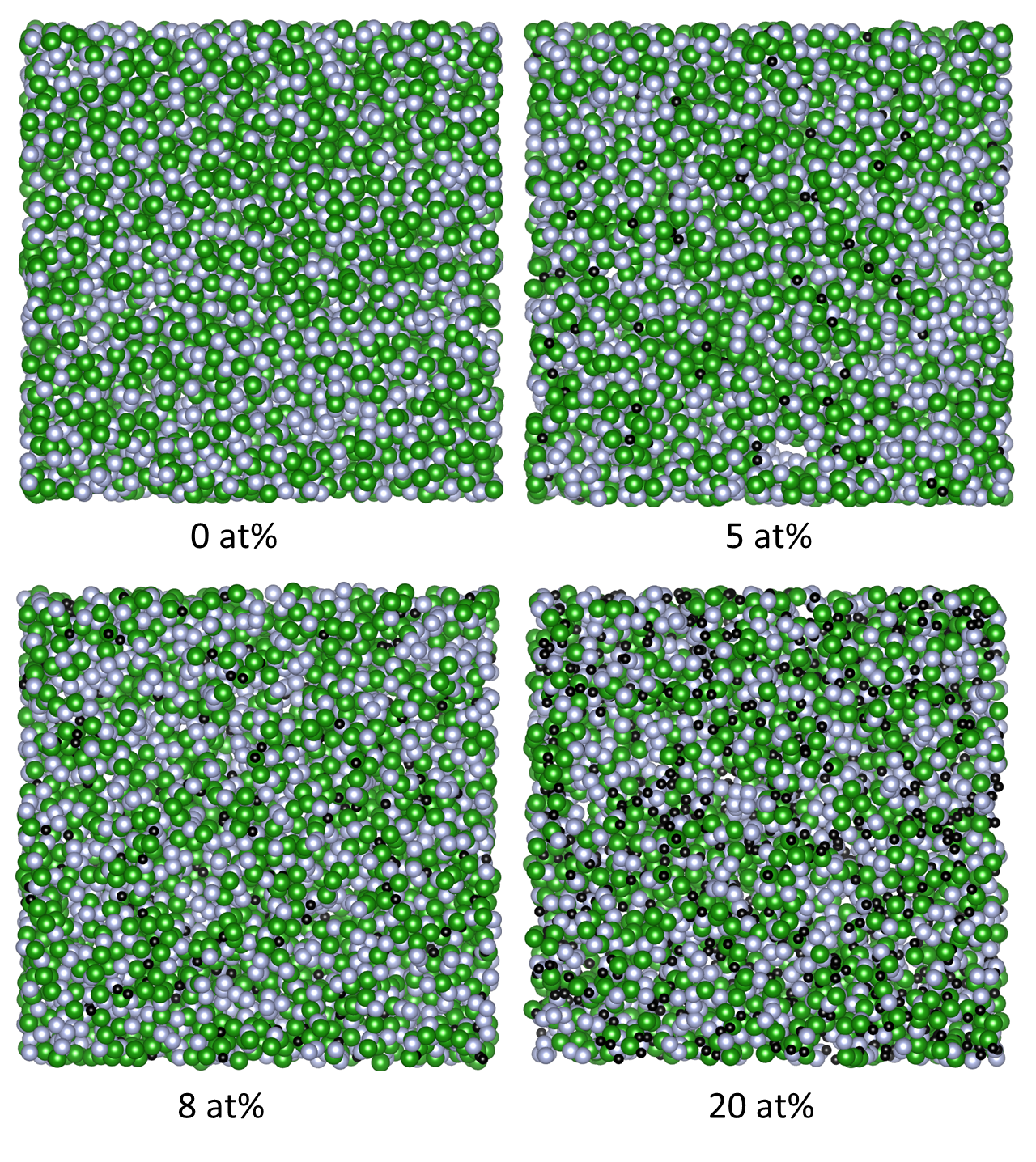}
    \caption{A subset of generated $\rm \alpha$-BN:H samples with varying H concentration using VESTA\cite{Momma2011} software where B, N, and H atoms are represented as black, grey, and black, respectively.}
    \label{fig:abnsamples}
\end{figure}

\begin{figure}[ht]
\centering
    \includegraphics[width=1.0\columnwidth]{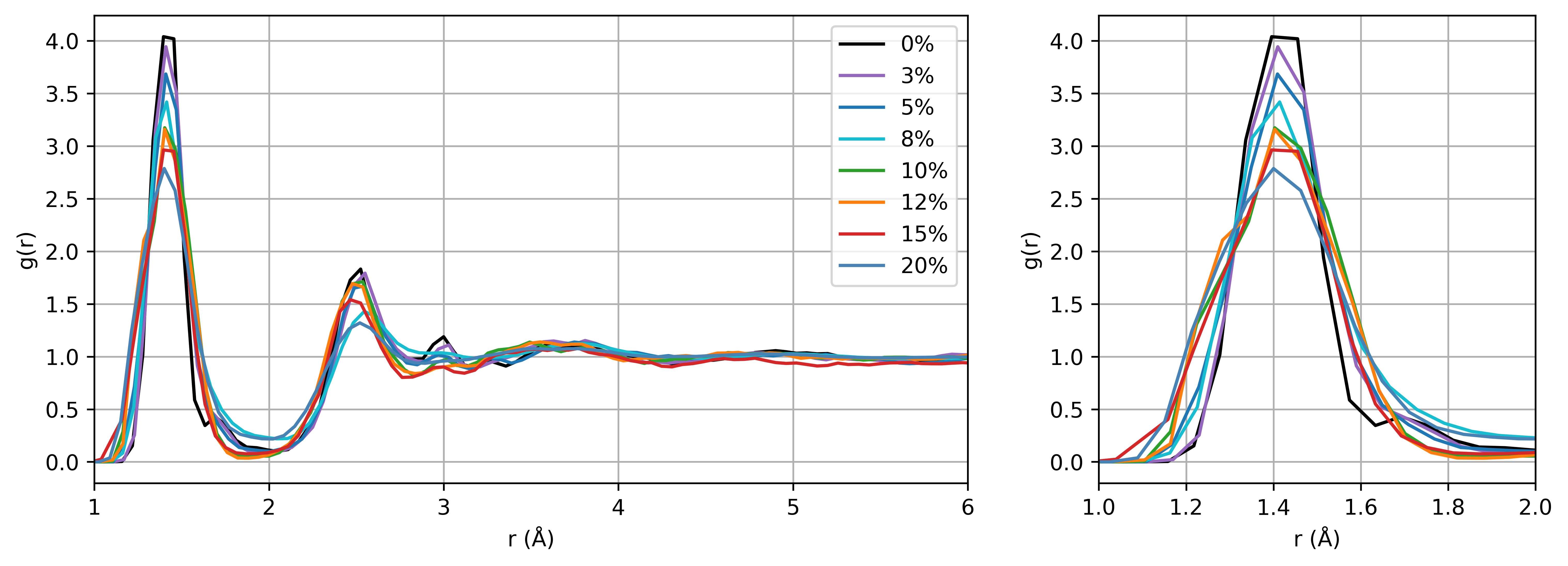}
    \caption{Average RDF of melt-quenched $\rm \alpha$-BN:H samples. All lines are averaged over five samples.}
    \label{fig:rdf}
\end{figure}

We first present the analysis of the morphology of $\rm \alpha$-BN:H with different H concentrations employing the melt-quench protocol. A subset of the samples generated in this work is shown in Fig. \ref{fig:abnsamples}. The RDF in Fig.~\ref{fig:rdf} shows the density of surrounding atoms as a function of distance and gives insight into the crystallinity of the material. Even though the first two peaks ($\leq$ 4 $\AA$) are clearly identified, no peak can be recognizable for longer distances, indicating the lack of long-range order. Hence, the amorphous character of $\rm \alpha$-BN does not change with the H doping.\\
The short-range order in $\rm \alpha$-BN:H is dominated by the first-neighbor distance, which contributes to the first peak in the RDF located at an average distance of $\sim$ 1.42 $\rm \AA$, as shown in Fig. ~\ref{fig:rdf}. A closer look at the first peak reveals an increased broadening and a shift to the left side with increased doping concentration, suggesting that it is induced by the presence of hydrogen atoms. Such change occurs due to the formation of chemical bonds different from B-N (B-H, N-H, B-B, and N-N), whose average bond length is different from B-N. The average bond lengths of these bonds are approximately $\sim$ 1.21 $\rm \AA$, $\sim$ 1.05 $\rm \AA$, $\sim$ 1.81 $\rm \AA$, and $\sim$ 1.44 $\rm \AA$, respectively.\\

\begin{figure}[ht]
\centering
    \includegraphics[width=1.0\columnwidth]{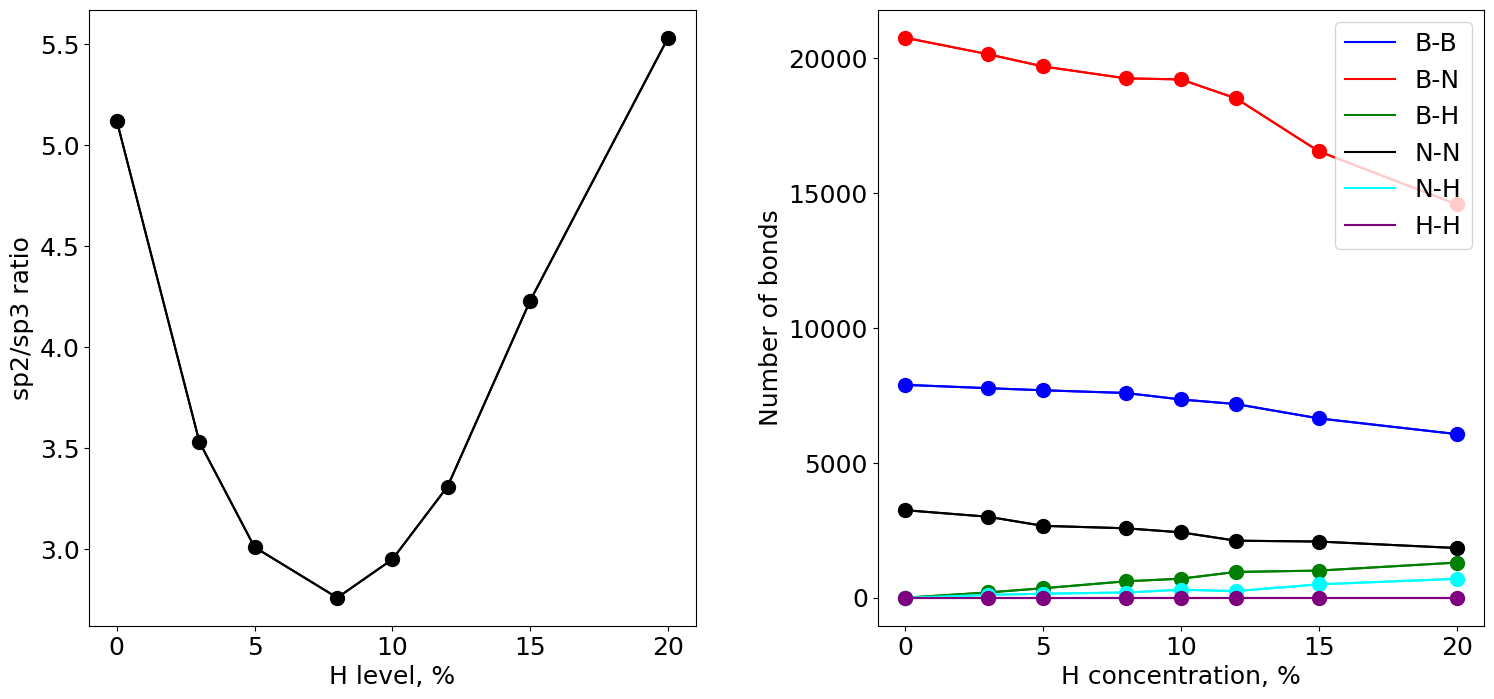}
    \caption{The ratio of $\rm sp^2$ and $\rm sp^3$-hybridized atoms of $\rm \alpha$-BN:H as a function of H concentration (left). Number of the chemical bonds with respect to the H concentration (right). }
    \label{fig:sp2}
\end{figure}

As $\rm \alpha$-BN:H is formed by the melt-quenching method from high temperatures, the resulting microstructure frozen in from the quenching process has a large influence in the hybridization of atoms in formed film. Understanding how hybridization can be changed with the fabrication conditions can help us tailor the properties of the material. The coordination number of $\rm \alpha$-BN:H is calculated to determine the type of hybridization. Fig.~\ref{fig:sp2} shows the ratio of $\rm sp^2$ (having coordination number 3) and $\rm sp^3$ (having coordination number 4) hybridized atoms. $\rm sp^1$-hybridized atoms (having a coordination number of 2) are also presented. With the introduction of H atoms to $\rm \alpha$-BN, the ratio of $\rm sp^2/sp^3$ drops rapidly with a minimum observed at 8 at\% H concentration. A deeper understanding of the chemical composition of the samples can be obtained by investigating the number and nature of chemical species involved in the bonds of the samples as a function of H concentration. As shown in Fig. ~\ref{fig:sp2}, while B-H and N-H bonds are increasing, others seem to be decreasing monotonically with larger H concentrations. No H-H bonds are observed up to 20 at\% of H concentration.\\
While having more H atoms reduces the total mass of the sample, canceling the $\rm sp^1$ hybridizations and having shorter bonds cause a dramatic shrinkage in the volume of the cell down to an H concentration of 8 at\%. Due to this interplay, the density of $\rm \alpha$-BN:H at low H concentration levels is increased from 2.17 to 2.181 $g/cm^3$, however, at larger H concentrations, the density of $\rm \alpha$-BN:H samples drops rapidly, as low as 2.01 $g/cm^3$.\\
In amorphous structures, such as $\rm \alpha$-BN and $\rm \alpha$-Si, some dangling bonds, vacancies and voids can be observed. H incorporation to $\rm \alpha$-Si can reduce the amount of dangling bonds and vacancies at small concentrations. Smets et al. \cite{Smets2003}  showed that when the H concentration is lower than 14 at\%, then H atoms bond with other Si atoms around Si divacancies. However, at higher concentrations, H atoms disrupt the Si network, reduce the density of $\rm \alpha$-Si:H structure and increase the size and number of voids within the structure \cite{Guerrero2020,  BEYER20122023, Young2007nanostructure}. Our simulations show a similar behavior for $\rm \alpha$-BN. While $\rm \alpha$-BN amount of $\rm sp^1$-hybridized atoms is reduced significantly with H incorporation lower than 8\%, some nanovoids occur and amount of $\rm sp^1$-hybridized atoms increase. Moreover, a sharp decrease in the amount of B-N bonds can be observed after H concentration of 12\%. This indicates that H atoms start to disrupt the B-N network and create these nanovoids, altering the mechanical properties and thermal stability of $\rm \alpha$-BN:H samples. \\

\subsection{Thermal Stability of $\rm \alpha$-BN:H}
\begin{figure}[ht]
\centering
    \includegraphics[width=\columnwidth]{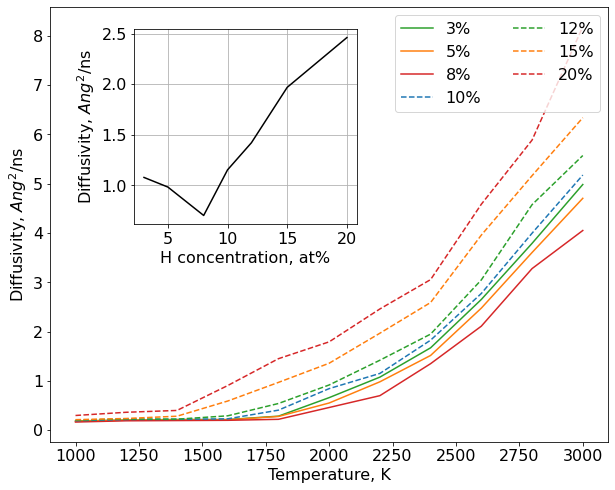}
    \caption{Diffusivity of H atoms as a function of temperature for samples of $\rm \alpha$-BN:H with varying H concentration. Inset: Diffusivity of H atoms at 2200 K as a function of level of H doping.}
    \label{fig:stab}
\end{figure}

Upon investigating the morphology of the generated samples, we also calculate the diffusivity of samples as a function of sample temperature. The diffusivity (as in described in Eq. 2) of samples at any given temperature can be extracted from the mean square displacement (MSD) of atoms where the MSD shows the average mobility of particles, as shown in Eq. 3 where \textbf{r} is position of the atom at any moment. The diffusivity of a sample is zero when MSD approaches a non-zero constant and has a zero slope. However, when the sample under investigation experiences a structural rearrangement and loses its stability, MSD has a non-zero slope. This allows us to evaluate the thermal stability of the samples and understand when they become unstable. Here, thermal stability refers to the material's ability to retain its structural integrity without significant atomic diffusion or rearrangement and loss of short-range order when subjected to high temperatures. In order to assess the thermal conductivity, we calculate the stability of samples between 300 K to 3000 K. We first calculate the MSD and diffusivity at that temperature for 70 ps and then increase the temperature of the samples by 50 K in 30 ps, later we calculate the MSD at the new temperature. The time intervals are determined to be large enough to obtain statistically meaningful data. The NPT (constant number of atoms, pressure, and temperature) ensemble with a Nose-Hoover thermostat has been applied with a timestep of 0.25 fs.\\
\begin{equation}
     \rm D=\lim _{t\rightarrow \infty}MSD(t)/6t
\end{equation}
\begin{equation}
    \rm MSD(t) = \langle |\textbf{r}_i(t)-\textbf{r}_i(0)|^2 \rangle
\end{equation}
The diffusivity of H atoms in $\rm \alpha$-BN:H is presented in Fig. \ref{fig:stab}. Non-zero values at low temperatures are obtained due to the vibration of atoms. Larger MSD values, which indicate structural rearrangement of atoms and unstable structure, are observed starting from 1600 K for 3 at\% and 5 at\% H-doping. For the case of 8 at\% doping, diffusivity is near zero up to 1800 K. For larger H doping this temperature value drops significantly. For highly H-doped $\rm \alpha$-BN, diffusivity becomes non-zero between 1000-1200 K. However, B and N atoms become diffusive at higher temperatures compared to H atoms. While H atoms begin to diffuse around 1800 K, B and N atoms still have near-zero diffusivity until 2000 K. Similarly at higher H concentration levels, there is a significant difference between the temperature at which H atoms and B and N atoms start to diffuse. At low-level H doping, this temperature difference is quite low.\\ 
Regardless of the identity of atoms, a monotonic trend between thermal stability and H concentration levels is observed. $\rm \alpha$-BN:H samples become more stable until the H concentration reaches 8 at\% due to the increase in $\rm sp^3$-hybridized atoms and reduction in $\rm sp^1$-hybridized atoms. Thereafter, thermal stability decreases rapidly since larger H doping reduces the density and causes some red nanovoids within the sample. At H concentrations larger than 20 at\%, samples become unstable. At low hydrogen levels, H doping can lead to a more thermally stable structure since it reduces the amount of $\rm sp^1$-hybridized B and N atoms, reduces the number of dangling bonds, and increases the density. However, at larger concentrations, H atoms disrupt the stability of the $\rm \alpha$-BN samples. Kaya et al. \cite{Kaya2023} showed that the thermal stability of C-doped $\rm \alpha$-BN samples can be improved with larger amounts of $\rm sp^3$-hybridized atoms. Similarly, Cloutier et al. \cite{CLOUTIER201465} and Liu et al. \cite{Liu2019} showed that increasing the density of $\rm \alpha$-C films and number of $\rm sp^3$-hybridized atoms can improve the stability of films.   \\

\subsection{Mechanical Properties of $\rm \alpha$-BN:H}

\begin{figure}[ht]
\centering
    \includegraphics[width=0.66\columnwidth]{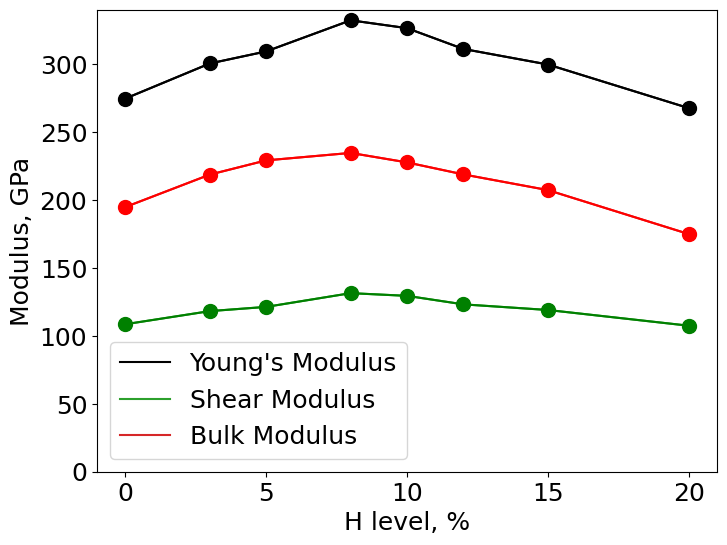}
    \caption{Mechanical properties of $\rm \alpha$-BN:H with respect to the H concentration.}
    \label{fig:mech}
\end{figure}

To calculate the mechanical properties of $\rm \alpha$-BN:H  samples as a function of H concentration, we compute the elastic constant of the samples by using the stress fluctuations and Born matrix (i.e., the second derivatives of energy with respect to the strain) in a NVT ensemble at 300 K\cite{Ray1984}. For each sample generated in this study, we calculate the full elastic stiffness tensor $C_{ij}$ using the LAMMPS. Later, we calculate Young's modulus, shear modulus, bulk modulus, and Poisson's ratio from the stiffness matrix.\\
Fig.~\ref{fig:mech} shows Young’s modulus, shear modulus, and bulk modulus of $\rm \alpha$-BN:H samples. Results show the same non-monotonic trend similar to thermal stability and $\rm sp^2/sp^3$ ratios. This clearly indicates that the mechanical properties are largely dependent on the microstructure of $\rm \alpha$-BN:H. Young’s modulus of pure $\rm \alpha$-BN samples was calculated as 270.11 GPa, which increases to 332.21 GPa with 8 at\% H concentration, respectively, which coincidentally have the largest $\rm sp^3$-hybridized bonds. The shear modulus and bulk modulus of $\rm \alpha$-BN samples are also increased with a higher density and larger number of $\rm sp^3$-hybridized atoms. However, larger H doping worsens the mechanical properties due to fewer $\rm sp^3$-hybridized atoms, lower density, and having more nanovoids. Reduction in mechanical properties with lower density and $\rm sp^3$-hybridized atoms has already been shown for $\rm \alpha$-BN and other amorphous structures \cite{Kaya2023, Lin2022, Bu2019, Lehmann2001, JACKSON1995195}. Even though the reported Young's modulus values in this study are lower than hexagonal and cubic BN, $\rm \alpha$-BN:H still has superior mechanical properties than other ultralow-dielectric materials. Another important mechanical property is Poisson's ratio, which gives us insight into how materials act under stress. Poisson's ratio of $\rm \alpha$-BN:H samples ranges between 0.24 and 0.281. Even though there is no clear trend with the Poisson's ratio and H doping or number of $\rm sp^3$-hybridized atoms, $\rm \alpha$-BN:H samples have a Poisson's ratio lower than 0.27, and Poisson's ratio drops significantly for structures with H doping higher than 15 at\%. Since materials with Poisson's ratio lower than 2/7 are assumed to be brittle\cite{Cao2016}, all $\rm \alpha$-BN:H samples are assumed to be brittle.\\
 \begin{figure}[ht]
\centering
    \includegraphics[width=1.0\columnwidth]{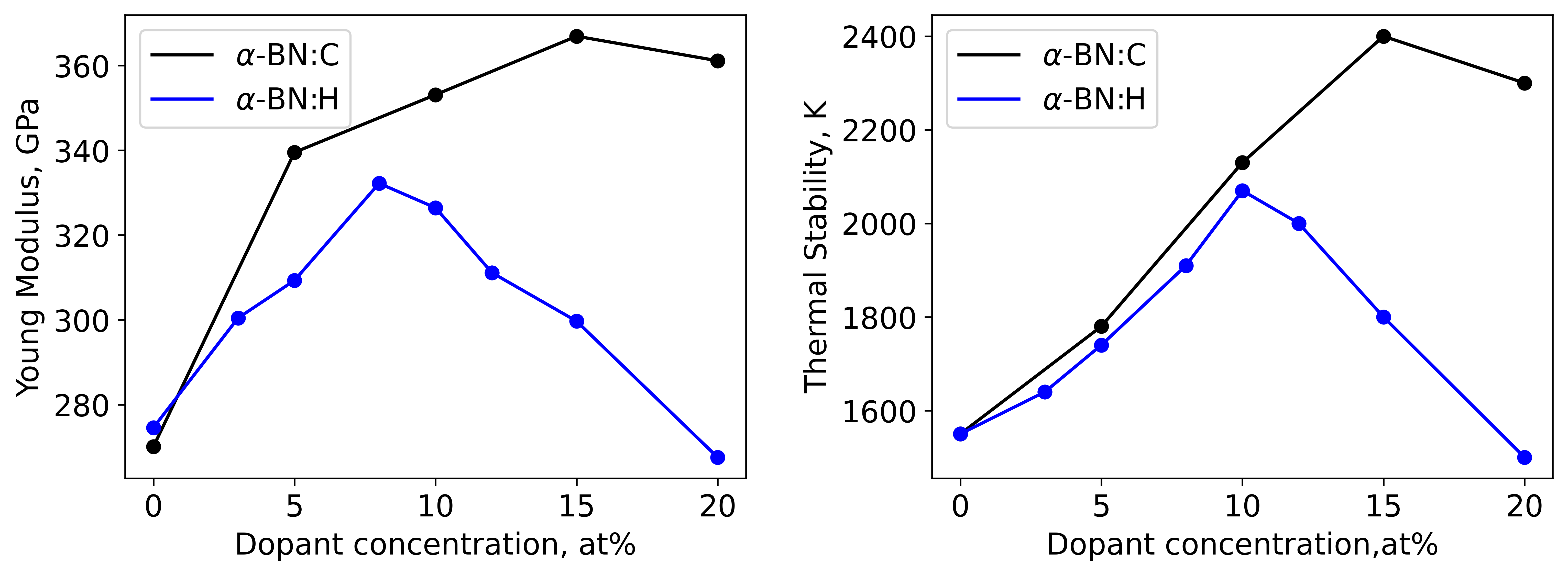}
    \caption{Comparison of the temperature that the $\rm \alpha$-BN:H and $\rm \alpha$-BN:C samples lose their stability and their Young's modulus with respect to the doping concentration.}
    \label{fig:comp}
\end{figure}  
The data in Fig. \ref{fig:comp} will allow us to get a deeper insight into how the microstructure of $\rm \alpha$-BN films changes with the doping, we compare the temperature at samples lose their thermal stability of B and N atoms, and Young's modulus of $\rm \alpha$-BN:H and $\rm \alpha$-BN:C reported by earlier\cite{Kaya2023}. The temperature values presented in the Fig. \ref{fig:comp} show the approximate temperature where the diffusivity of B and N atoms becomes a non-zero value. Since C atoms can bond with four atoms instead of one and are heavier than H atoms, they lead to more $\rm sp^3$ bonds and denser samples than $\rm \alpha$-BN:H. This leads to a higher Young's modulus and more stable structures, even at higher doping values (20 at\%). Variation in stability and mechanical properties due to doping shows that the fundamental properties of $\rm \alpha$-BN films can be altered in different fabrication conditions.  \\

\section{Conclusion}
Excellent mechanical properties and ultralow dielectric constant of $\rm \alpha$-BN opens a new pathway for microelectronics and neuromorphic computing technologies. This study reveals how H doping tunes the morphology, stability, and mechanical properties of $\rm \alpha$-BN. We first develop a machine-learning interatomic potential for H dopant in $\rm \alpha$-BN and performed GAP-driven MD simulations to generate realistic structures. Thanks to the accurate machine learning approaches, we show that thermal stability and mechanical properties of $\rm \alpha$-BN are improved by small H doping levels. Despite $\rm \alpha$-BN:H's extraordinary properties, it is crucial to perform a thorough benchmark analysis on growth conditions and corresponding properties. The results obtained in our study will provide a guide to process engineers to optimize the growth conditions to achieve the optimum materials performance in the context of microelectronics.

\section*{Acknowledgement}
This project has been supported by Samsung Advanced Institute of Technology and is conducted under the REDI Program, a project that has received funding from the European Union's Horizon 2020 research and innovation programme under the Marie Skłodowska-Curie grant agreement no. 101034328. This paper reflects only the author's view and the Research Executive Agency is not responsible for any use that may be made of the information it contains. ICN2 acknowledges the Grant PCI2021-122092-2A funded by MCIN/AEI/10.13039/501100011033 and by the “European Union NextGenerationEU/PRTR”. Simulations were performed at the Center for Nanoscale Materials, a U.S. Department of Energy Office of Science User Facility, supported by the U.S. DOE, Office of Basic Energy Sciences, under Contract No. DE-AC02-06CH11357. Additional computational support was received from the King Abdullah University of Science and Technology-KAUST (Supercomputer Shaheen II Cray XC40) and Texas Advanced Computing Center (TACC) at The University of Texas at Austin.

\section*{References}
\bibliographystyle{iopart-num}
\bibliography{bibliography}

\end{document}